\documentclass[5p,times,sort&compress]{elsarticle}
\usepackage{amsmath,amssymb}
\usepackage{graphicx}
\usepackage{siunitx}
\usepackage{color}
\usepackage{soul} 

\begin{document}

\title{Fluid Vesicles in Flow}
\author[stut]{David Abreu\corref{cor1}}
\ead{david.abreu@itp2.uni-stuttgart.de}
\author[weiz]{Michael Levant}
\ead{michael.levant@weizmann.ac.il}
\author[weiz]{Victor Steinberg}
\ead{victor.steinberg@weizmann.ac.il}
\author[stut]{Udo Seifert}
\ead{useifert@theo2.physik.uni-stuttgart.de}
\cortext[cor1]{Corresponding author}
\address[stut]{{II.} Institut f\"ur Theoretische Physik, Universit\"at Stuttgart, 70550 Stuttgart, Germany}
\address[weiz]{Department of Physics of Complex Systems, Weizmann Institute of Science, Rehovot, 76100, Israel}

\begin{abstract}
We review the dynamical behavior of giant fluid vesicles in various types of external hydrodynamic flow. The interplay between stresses arising from membrane elasticity, hydrodynamic flows, and the ever present thermal fluctuations leads to a rich phenomenology. In linear flows with both rotational and elongational components, the properties of the tank-treading and tumbling motions are now well described by theoretical and numerical models. At the transition between these two regimes, strong shape deformations and amplification of thermal fluctuations generate a new regime called trembling. In this regime, the vesicle orientation oscillates quasi-periodically around the flow direction while asymmetric deformations occur. For strong enough flows, small-wavelength deformations like wrinkles are observed, similar to what happens in a suddenly reversed elongational flow. In steady elongational flow, vesicles with large excess areas deform into dumbbells at large flow rates and pearling occurs for even stronger flows. In capillary flows with parabolic flow profile, single vesicles migrate towards the center of the channel, where they adopt symmetric shapes, for two reasons. First, walls exert a hydrodynamic lift force which pushes them away. Second, shear stresses are minimal at the tip of the flow. However, symmetry is broken for vesicles with large excess areas, which flow off-center and deform asymmetrically. In suspensions, hydrodynamic interactions between vesicles add up to these two effects, making it challenging to deduce rheological properties from the dynamics of individual vesicles. Further investigations of vesicles and similar objects and their suspensions in steady or time-dependent flow will shed light on phenomena such as blood flow.  
\end{abstract}


\maketitle 
  
\begin{section}{Introduction}
  Giant unilamellar vesicles (GUVs) have become a paradigmatic soft matter system for many reasons. First, even in equilibrium they exhibit an amazing variety of shapes. At fixed area and enclosed volume, these shapes result from the minimization of the bending, or "curvature", energy 
    \begin{equation}
      {\cal H}=\int dA\left[\frac{\kappa}{2}(2H-C_0)^2+\kappa_G K\right]
      \label{eq_bend}
    \end{equation}
  written here in the form introduced by Helfrich forty years ago \cite{helf73,deul76}. $H$ and $K$ are the mean and Gaussian curvatures, respectively, $\kappa$ and $\kappa_G$ the corresponding bending energies, and $C_0$ a spontaneous curvature reflecting bilayer asymmetry. Systematic theoretical work in the nineties in combination with experiments using video microscopy has led to a quantitative understanding of how the subtle aspects of bilayer elasticity \cite{shee74,evan74,svet89} determine the shape diagram, as comprehensively reviewed in \cite{seif97}. Second, even though biological cells have a more complex architecture, vesicles have often served as a model system for anucleate cells such as red blood cells (RBCs), whose equilibrium shapes can be predicted \cite{lim02,khai11} by minimizing a generalized version of the energy \eqref{eq_bend}. Third, and coming to the topic of this review, the behavior of vesicles in external flow field is determined by a complex interplay between membrane elasticity, hydrodynamic forces, and thermal fluctuations acting at microscopic length scales. Studying the resulting rich phenomenology is fundamental for understanding the flow dynamics of these soft objects.

  The physical properties of the membrane play a key role in this dynamics. The lipid bilayer with a thickness of approximately \SI{5}{\nano\meter} is very small compared to the GUV radius ($\sim$\SI{10}{\micro\meter}). This bilayer is often in a liquid phase at room temperature \cite{dimo06} making the vesicle very deformable. Bending deformations involve much lower energies than stretching and shearing ones which can be neglected \cite{seif97}. The bending rigidity $\kappa$ is typically between $10^{-20}$ and \SI{2d-19}{\joule} (approximately $2-50 k_BT$ at room temperature, although lipids widely used in dynamical experiments have a much narrower range of $25-50 k_BT$) \cite{seif97,rawi00,dimo06,geno13}. The membrane viscosity is about $10^{-8}$ to \SI{d-9}{\pascal\second\meter} \cite{waug82a,rawi00,dimo06,petr12,wood12,hone13}. Typically, the membrane can be considered incompressible since the number of lipids in it is constant and the stretching energy is very large \cite{seif99}. Therefore, the total membrane area is constant. Moreover, the membrane is permeable to water but impermeable to many other molecules. For a vesicle in equilibrium, any influx of water creates an osmotic pressure which is relaxed by an outflux of the same magnitude \cite{helf73}. In experiment, one tries to keep a zero net osmotic pressure. Even if a small net flow still exists, volume changes occur on a time-scale of several hours, which is much longer than the typical experimental time-scale of about 10 to 15 minutes. We can thus consider the vesicle volume to be constant as well. These few properties are sufficient to characterize the forces that will counteract the external forcing.
  
  We first review the general theoretical, numerical, and experimental methods used to address this problem. Then, we consider the case of planar linear flow, for which vesicle dynamics has been thoroughly studied in recent years. Afterwards, we describe the effect of walls and capillary flows on single vesicles. Moving further up the scale from micro to macro, we discuss hydrodynamic interactions and the rheology of vesicle suspensions. Finally, we present related questions on the non-equilibrium dynamics of vesicles and similar objects. 
\end{section}

\begin{section}{Methods}
  We present the theoretical tools for describing the dynamics of vesicles under hydrodynamic flows, then the techniques used in direct numerical simulations, and finally the experimental setups.

  \begin{subsection}{Theoretical modeling}
  \label{theor}
    The membrane is modeled as a two-dimensional sheet of incompressible fluid \cite{powe10}. It encloses an internal liquid of viscosity $\eta_i$ and is suspended in an outer liquid of viscosity $\eta_o$, defining the viscosity contrast $\lambda\equiv\eta_i/\eta_o$. The volume $V$ and the surface area $A$ are constant but the vesicle is not necessarily spherical. We thus define the effective radius $R_0\equiv (3V/4\pi)^{1/3}$, which is the radius of a sphere of the same volume. Relatively to this sphere, the vesicle has an excess area \footnote{Some authors define an effective radius over the area as $R'_0\equiv\sqrt{A/4\pi}$ and a reduced volume $\tau\equiv V/(4\pi {R'_0}^3/3)\leq1$. The excess area is then given by $\Delta=4\pi(\tau^{-2/3}-1)$.} $\Delta\equiv A/R_0^2 -4\pi\geq0$. Ignoring for simplicity the energy due to Gaussian curvature (constant for spherical geometry) and spontaneous curvature, the bending energy of the membrane is given by \cite{canh70,helf73}
	  \begin{equation}
	    {\cal H}_\kappa=\int dA\frac{\kappa}{2}\left[(2H)^2+\sigma\right],
	    \label{helf}
	  \end{equation}
    where $\sigma$ is the surface tension, a Lagrange multiplier that ensures local and global area conservation. Unlike for droplets, $\sigma$ is here a dynamical variable, analogous to pressure for three-dimensional fluids, which adjusts itself to compensate the external stresses. It can therefore take negative values as explained further below. This vesicle is subject to an external flow. Due to the vesicle dimensions, the Reynolds number $Re$ is small - for a vesicle of radius \SI{10}{\micro\meter} suspended in water (viscosity of \SI{d-3}{\pascal\second}) and subject to a shear flow with rate \SI{1}{s^{-1}}, $Re \sim 10^{-4}$ - and the flow is described by the Stokes equations
	  \begin{align}
	    {\bf \nabla v}=0& &\nabla p=\eta{\bf \nabla}^2 {\bf v},
	  \end{align}
    where $\bf v$ is the flow velocity, $p$ the pressure, and $\eta$ the viscosity. These equations have to be solved for the inner and outer fluids. The velocities and stresses are then matched at the membrane with no-slip boundary conditions and under the constraints of membrane incompressibility and impermeability.

    \begin{figure}
      \centering
      \includegraphics[width=0.7\linewidth]{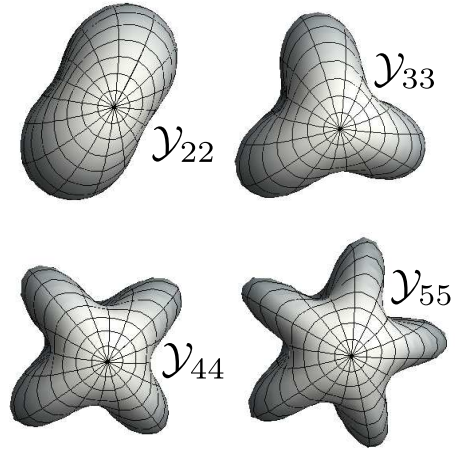}
      \caption{Examples of spherical harmonics. The represented function is $r(\theta,\phi)=1+\text{Re}({{\cal Y}_{ll}(\theta,\phi)})$ for $l=2,3,4,5$.}
      \label{fig_harm}
    \end{figure}
    Analytical models need further assumptions to derive equations of motion for the vesicle. One strategy consists in describing the dynamics effectively with only a few degrees of freedom \cite{kell82,seco82,riou04,nogu05,made07,nogu07,abre12}. These models are based on the Keller-Skalak (KS) model \cite{kell82}, which assumes vesicles of fixed ellipsoidal shape with fluid membrane. Their dynamics in shear flow is then described by only two variables: the inclination angle $\theta$ of the long axis of the vesicle relative to the flow direction, and the angle $\phi$ describing the displacement of a membrane element, see Fig. \ref{fig_tttb}. Another strategy relies on looking at quasi-spherical vesicles, i.e., vesicles with small excess area $\Delta\ll1$ \cite{youh95,olla97,seif99,roch05,misb06,lebe07,vlah07,abre13}. The radius of such vesicles
      \begin{equation}
	r(\theta,\phi)\equiv R_0\left(1+\sum_{l,m}u_{lm}{\cal Y}_{lm}\right)
	\label{eq_quasis}
      \end{equation}
    can be expanded in spherical harmonics ${\cal Y}_{lm}$, see Fig. \ref{fig_harm}. Then, the inner and outer pressure and velocity fields are given by Lamb's solution of the Stokes equations \cite{happ83}. The main approximation is to match these fields not at the membrane, but rather at the virtual sphere of radius $R_0$ \cite{seif97}. This scheme can be applied in a perturbative way to analyze larger distortions from the spherical shape \cite{dank07a,lebe07,faru10}. Moreover, thermal fluctuations can then be easily included \cite{seif99,fren03,fink08,abre13}. Of course, the equations obtained in such a way often have to be solved numerically \cite{nogu07,turi08,me09,faru10,nogu10c,abre12,abre13}. However, we will distinguish numerical solutions of perturbative equations from direct numerical simulations as discussed in the next section. 
  \end{subsection}

  \begin{subsection}{Numerical approaches}
    Analytical models are either effective models or apply only in the quasi-spherical limit. Therefore, one has often to resort to direct numerical simulations to make quantitative predictions that can be compared to experiments. These numerical models either treat the fluids and the membrane on a continuum-level or model parts of the system as effective particles (see the review \cite{li13} for RBCs).
    
    Continuum-based models solve the hydrodynamic equations directly. The boundary-integral method \cite{pozr92,krau96,cant99,cant99a,suku01,kaou08,coup08,me09,veer09,bibe11,boed11,kaou11a,veer11,zhao11,zhao11b,faru12a,faru12b,faru13,liu13,zhao13} uses Green's tensor formalism to calculate velocity fields. These are deduced by integration of the membrane forces over the vesicle surface. In this scheme, one does not need to follow explicitly the boundary of the vesicle. Moreover, the calculated velocity field obeys the Stokes equations per definition. Recently, spectral methods \cite{dimi07,kess08,veer09,zhao10,zhao11} have increased the accuracy of such simulations. However, the boundary-integral method is limited to viscous flows and cannot include inertial effects. Other continuum-based methods offer more flexibility for including inertial and non-linear effects. Immersed boundary algorithms \cite{pesk02,taka09,kim10,kim12} use the similarity of hydrodynamics and elasticity equations to simulate incompressible viscoelastic bodies immersed in incompressible fluids. The advected- or phase-field method \cite{bibe03,beau04,bibe05} replaces the membrane by a continuous field which varies strongly at the boundary between inner and outer fluids. Another continuum-based numeric schemes include level-set \cite{laad12,sala12,doye13} and front-tracking \cite{yazd12,luo13} methods. One important drawback of continuum-based models is that they fail to take thermal fluctuations into account, since methods existing for flat membranes \cite{brow08} have not been yet generalized to vesicles. This deficiency, which may seem negligible since $\kappa/k_BT\gtrsim25$, explains why such models fail to predict the trembling motion observed experimentally, see Section \ref{sec_tr}.

    Particle-based methods define effective particles that interact in such a way that the actual equations of motion are recovered. Lattice-Boltzmann algorithms \cite{kaou11,kaou12,kaou13,hall13} discretize the fluids while keeping a continuum description of the membrane. These are, however, limited to two-dimensional models without thermal fluctuations. In the stochastic Euler-Lagrangian method \cite{bran13}, the equations of fluctuating hydrodynamics are solved while the membrane is coarse-grained. Dissipative particle dynamics can also include thermal fluctuations \cite{sugi07}. However, both methods have not been yet applied to vesicle dynamics. More interestingly, the multi-particle collision dynamics method \cite{nogu05,nogu05a,nogu05b,nogu07,fink08,mcwh09,me09,nogu10c,nogu10d,lamu13} takes hydrodynamic interactions, membrane properties, and thermal fluctuations into account. This numerical scheme is very flexible and allows to simulate complex flow geometries and many-body problems.
  \end{subsection}

  \begin{subsection}{Experimental setups}
    Most experiments use dioleyol-phosphatidylcholine (DOPC) vesicles whose membrane is fluid at room temperature \cite{abka02a}. There are a lot of methods to produce them \cite{wald10}. Due to their size, GUVs can be directly observed with phase-contrast microscopes (e.g. \cite{abka02a}). Epifluorescent microscopy is used to better detect and analyze the vesicle shape \cite{kant05}. There, fluorescent lipids are added to the vesicle membrane during its formation. Both methods allow only to observe a two-dimensional cut of the three-dimensional vesicle. Three-dimensional images can be obtained either by confocal laser scanning microscopy \cite{lorz00,saka12,hone13} or by mechanical scanning of the narrow focal plane \cite{desc09}. However, such scanning cannot be applied to monitor fast dynamical changes such as local shape perturbations. Markers can be used to track membrane or fluid motion, for example when the vesicle tank-treads \cite{abka02a,kant05,hata11,hone13}.

    Several methods exist to generate an external flow. The most straightforward one is to generate a laminar flow in a micro-channel of rectangular section. This flow is parabolic but can be approximated to a linear shear close to the walls. Depending on the experimental question, vesicles can be exposed to a pure shear flow \cite{lorz00,razp00,fa04,abka05,kant05,kant06,made06,vezy07,kant08,hone13,pomm13} or to a Poiseuille flow \cite{vitk04,coup08,coup12} by properly adjusting the size of the micro-channel and the position of the vesicle in it. The channel can also be structured to investigate more complex geometries \cite{nogu10d,brau11}. One significant drawback of this method is that vesicles can only be observed for short periods since they flow. Observation periods can be slightly extended by placing a vesicle at the center of a shear flow with zero mean velocity, either by moving two parallel walls \cite{desc09} or two rotating cylinders \cite{haas97,shah98,call08,hata11} in opposite directions with the same velocity. However, such setups introduce additional experimental difficulties since vesicles are positioned in an unstable location and tend to escape. The same remark applies to cross-slot micro-channels, which consist in two opposite inputs perpendicular to two opposite outputs, thus generating pure elongational flow \cite{kant07,kant08a}. These stability limitations can be coped with by means of a closed-loop feedback system which corrects the flow. A significant improvement in observation periods is achieved in a micro-fluidic realization of the four-roll mill setup \cite{lee07,desc09a,zabu11,leva12,leva12a}. Such a setup allows to trap vesicles for observation periods significantly larger than the characteristic time-scales of their dynamics and generates a well-controlled general linear flow, see Fig. \ref{fig_flow}.
  \end{subsection}
\end{section}

\begin{section}{Dynamics in linear flows}
\label{sect_lin}
  \begin{figure}
    \centering
    \includegraphics[width=0.9\linewidth]{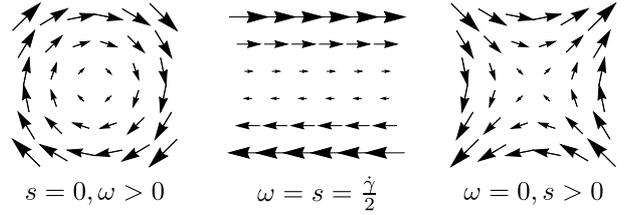}
    \caption{A general linear flow is the sum of a rotational (left) and an elongational component (right). A shear flow (middle) corresponds to both components having the same magnitude.}
    \label{fig_flow}
  \end{figure}
  In this section, we consider unbounded planar linear flows, for which the unperturbed velocity profile in Cartesian coordinates has the form
    \begin{equation}
      {\bf v}^\infty \equiv \omega(y{\bf e}_x-x{\bf e}_y)+s(y{\bf e}_x+x{\bf e}_y)
      \label{linflow}
    \end{equation}
  in each plane with constant $z$, where ${\bf e_x}$ and  ${\bf e_y}$ are the basis vectors of the $(xy)$-plane. The elongational flow of strength $s$ tends to stretch the vesicle along the (y=x)-direction, while the rotational flow of vorticity $\omega$ causes rigid-body-like rotation. Varying $s$ and $\omega$ leads to different types of flow; for example, setting $\omega=s\equiv\dot\gamma/2$ gives a linear shear flow with shear rate $\dot\gamma$, see Fig. \ref{fig_flow}. Apart from the excess area $\Delta$ and the viscosity contrast $\lambda$ defined in Section \ref{theor}, a third parameter, the capillary number
  \begin{equation}
    Ca\equiv \dot{\gamma}\eta_o R_0^3/\kappa
  \end{equation}
  will be important for the dynamics of a vesicle in linear flow. This dimensionless quantity compares the time-scale $1/\dot\gamma$ of the external flow with the relaxation time $\eta_oR_0^3/\kappa$ of bending deformations, where $\eta_o$ is the external viscosity and $R_0$ the effective radius. The capillary number thus quantifies the relative strength of hydrodynamic forces to bending ones. The characterization of the vesicle dynamics is in most cases done by looking at the projection of the vesicle shape in middle flow plane, that is, the plane with constant $z$ containing the center of mass of the vesicle. Of course, due to the constraints of constant volume and area, three-dimensional deformations occur if the vesicle deforms in the $(xy)$-plane but the dynamics is mostly characterized by what happens in this middle flow plane.

  \begin{subsection}{Tank-treading and tumbling}
    \begin{figure}
      \centering
      \includegraphics[width=0.9\linewidth]{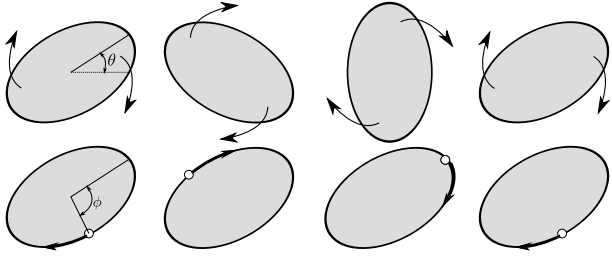}
      \caption{Tumbling (top) and tank-treading (bottom) motions of an ellipsoidal vesicle. $\theta$ is the inclination angle of the long axis of the vesicle, while $\phi$ measures the displacement of a membrane element relative to this orientation.}
      \label{fig_tttb}
    \end{figure}
    At low capillary numbers $Ca\ll1$, the deformation of the vesicle remains small. The KS model \cite{kell82} describes the two possible dynamical regimes, which were first observed in experiments with RBCs \cite{schm69,gold72,fisc78,basu11}. In the tumbling (TB) regime, the vesicle rotates as a whole like a rigid body, see Fig. \ref{fig_tttb}. In the tank-treading (TT) regime, the vesicle adopts a constant orientation in the flow plane (up to thermal fluctuations) while its membrane rotates around the interior fluid. The TT motion is a consequence of the membrane fluidity. In the KS model, TT happens below and TB above a critical viscosity contrast $\lambda_c(\Delta)\geq 1$, which is independent of the strength of the shear flow but decreases with increasing excess area $\Delta$ \cite{faru12a}. Whereas this does not hold for RBCs \cite{abka07,skot07,kess08,dupi12}, for which this model was originally conceived, it describes remarkably well the TT-TB transition for fluid vesicles at low shear rates. The TT-TB transition happens at lower $\lambda_c$ if the membrane viscosity is increased \cite{nogu04} since it contributes to an effective increase of the inner viscosity. Taking into account a possible slip between the two layers of the membrane also changes the critical viscosity contrast \cite{schw10}. If we consider a general linear flow, the TT-TB transition can be triggered at fixed viscosity contrast by increasing the relative strength $\omega/s$ of rotational to elongational flow. Therefore, contrary to experiments in pure shear flow, in which the viscosity contrast determines the dynamics, a four-roll mill experiment \cite{desc09a,leva12a} allows to explore different dynamical regimes with a single vesicle. We will, however, focus on the case of a pure shear flow to describe TB and TT.

    The inclination angle $\theta$ of the vesicle evolves in the KS model according to
    \begin{equation}
      \partial_t \theta=-\frac{\dot{\gamma}}{2}\left(1-B\cos(2\theta)\right),
      \label{thetatb}
    \end{equation}
    where $B>1$ is a constant depending on the vesicle geometry \cite{kell82}. This equation is obtained by first balancing the torques coming from the TT motion, the TB motion, and the external shear flow, and  second by equating the work applied by the flow on the membrane with the energy dissipated inside the vesicle. This equation describes relatively well the experimental TB motion \cite{razp00,kant06} and numerical results \cite{beau04,bibe11}, at least for small capillary numbers. For larger ones, corrections due to deformations are needed \cite{made06,bibe11}. Upon approaching the TB-TT transition, a critical slowing down of the TB frequency is observed \cite{kant06,lebe07} because the transition is a saddle-node bifurcation. Models including thermal fluctuations show additional dynamical features \cite{nogu05,abre12} which are supported to some extent by experimental observations \cite{kant05,kant06,zabu11}.
    
    In the TT regime, models and experiments have analyzed mainly three characteristics: membrane velocity, vesicle deformation, and inclination angle. The membrane velocity can be measured by tracking markers on the membrane \cite{kant06,vezy07} or in the fluid \cite{hata11}. The membrane velocity is periodic in $\phi$ (defined in Fig. \ref{fig_tttb}) \cite{kant06,vezy07}, in accordance with the KS model which predicts a constant rotation frequency $\partial_t\phi$. Moreover, this rotation frequency is proportional to the shear rate \cite{krau96,vezy07}. The KS model, however, does not allow for deformations. The vesicle deforms into an ellipsoid under the action of the elongational part of the flow. This deformation saturates at large shear rates because the excess area is limited and the membrane cannot be stretched. For quasi-spherical vesicles, the Taylor parameter $D\equiv(L-S)/(L+S)$, with $L$ and $S$ being the lengths of the long and short axes, saturates at a value proportional to $\sqrt{\Delta}$ \cite{seif99} in good agreement with experimental results \cite{haas97,kant05}. Finally, the inclination angle decreases with increasing viscosity contrast $\lambda$ and increasing excess area $\Delta$ \cite{kant05,kant06}. Its value and the fluctuations around it are very well described by the analytical theory for quasi-spherical vesicles \cite{seif99}. It is also noteworthy that recent direct numerical simulations \cite{zhao11,yazd12,bibe11} show an excellent agreement with experiments in the TT as well as in the TB regime. Recently, the TT motion has been analyzed as a means of propulsion in shear flow if the membrane bending rigidity varies locally \cite{olla11}.
  \end{subsection}

  \begin{subsection}{Deformations and transition regime}
  \label{sec_tr}
    \begin{figure}
      \centering
      \includegraphics[width=0.9\linewidth]{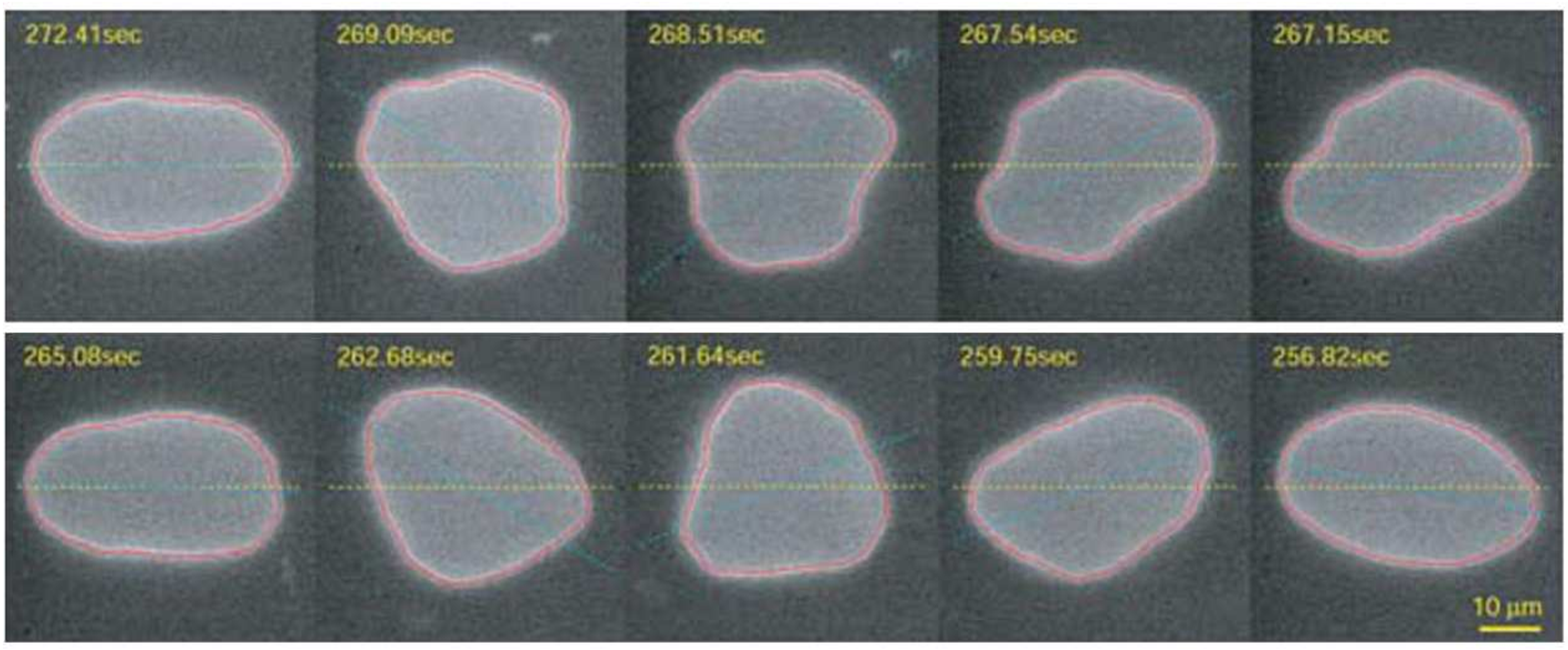}
      \includegraphics[width=0.9\linewidth]{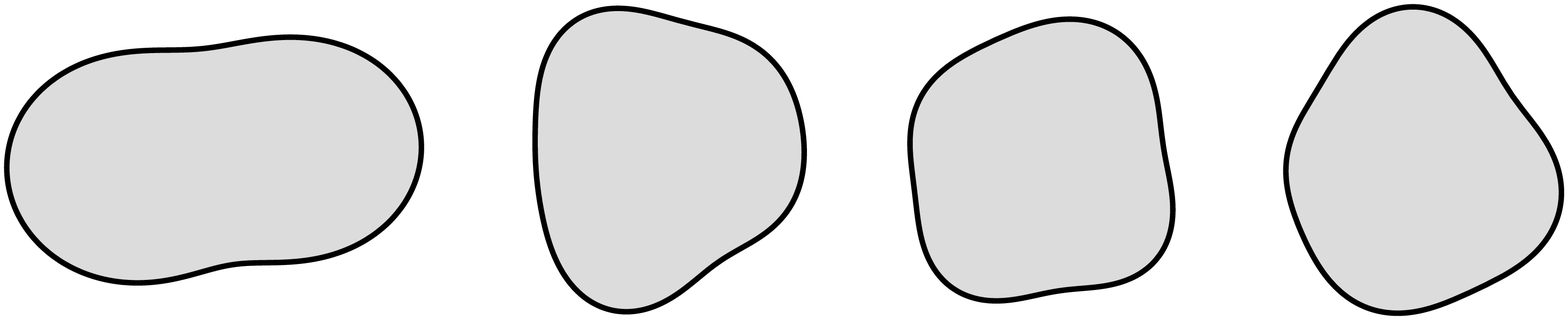}
      \caption{Top: Experimental snapshots of two TR pseudo-periods (from \cite{leva12}). Bottom: Shapes obtained in simulations with thermal noise at the same physical parameters (from \cite{abre13}).}
      \label{fig_trem}
    \end{figure}
    For larger capillary numbers $Ca\sim1$, vesicle deformation changes the dynamics. A new regime appears at the TT-TB transition which was first observed experimentally in \cite{kant06}. This transition regime called trembling (TR) is characterized by irregular oscillations of the vesicle orientation accompanied by strong and asymmetric shape perturbations \cite{kant06,desc09,desc09a,zabu11,leva12a}. In contrast, analytical models without thermal fluctuations predict a transition motion called vacillating-breathing (VB) in which the orientation angle oscillates around $0$ while the shape deforms with the same period \cite{misb06,vlah07,lebe07,dank07,dank07a,lebe08,gued12}. These models, based on the quasi-spherical expansion Eq. \eqref{eq_quasis}, consider only harmonics with $l=2$ since they possess the $r\rightarrow-r$ symmetry of the flow and are the only ones excited to lowest order. Refined models including higher-order even harmonics \cite{faru10,zhao11} and deterministic numerical simulations \cite{bibe11,zhao11,yazd12} also predict only periodic and symmetrical motions.

    These models ignore thermal noise because bending forces are much larger than thermal ones since $\kappa/k_BT\gtrsim25$. However, a detailed experimental examination has shown an amplification of thermal noise in TR, which is characteristic of non-linear dynamical systems close to bifurcations such as the TT-TB one \cite{leva12a}. In particular, the aperiodicity of the orientation angle and the presence of odd harmonics \cite{zabu11,leva12a} cannot be explained by deterministic predictions \cite{faru10,bibe11,zhao11,yazd12}.\footnote{The presence of odd harmonics in the deterministic simulations of Ref. \cite{yazd12} is, as the authors admit, only due to numerical errors so that they do not observe the experimental TR. However, for shear rates much larger than those used in experiments on TR, they observe higher-order even harmonics which are a feature of the wrinkling instability explained in Section \ref{sec_el2}} On the other hand, simulations and phenomenological models with thermal fluctuations show similarities with the experimental TR motion \cite{nogu04,nogu05,nogu07,me09}. A recent stochastic model based on a quasi-spherical expansion reproduces the characteristics of the TR motion, as exemplified in Fig. \ref{fig_trem} \cite{abre13} . Therefore, thermal fluctuations are most probably the reason for the discrepancy between the predicted VB motion and the observed TR one.
  \end{subsection}

  \begin{subsection}{Phase diagram}
    \begin{figure}
      \centering
      \includegraphics[width=0.7\linewidth]{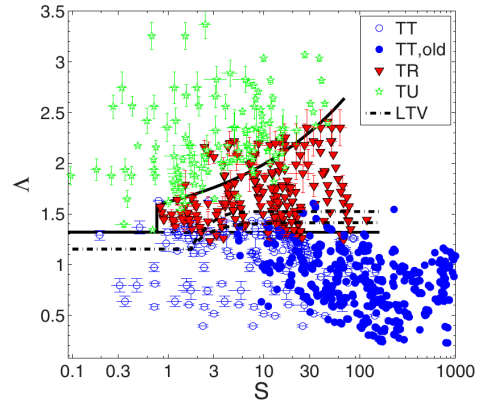}
      \caption{Experimental phase diagram of a vesicle in linear flow as a function of the reduced parameters $\Lambda$ and $S$ (from \cite{zabu11}). Blue and white dots correspond to TT motion, red triangles to TR and green stars to TU. The dash-dotted lines are the theoretical prediction from \cite{lebe08} and the solid ones are a guide for the eye.}
      \label{fig_pd}
    \end{figure}
    The theoretical model of Refs. \cite{lebe07,lebe08}, based on a quasi-spherical expansion to lowest order and including next-order corrections for the bending energy only, predicts that the phase diagram of the system merely depends on the two parameters
      \begin{align}
	S\equiv\frac{14\pi}{3\sqrt{3}}\frac{\eta_oR_0^3}{\kappa\Delta}s & &\Lambda\equiv\frac{23\lambda+32}{24}\sqrt{\frac{3\Delta}{10\pi}}\frac{\omega}{s}.
      \end{align}
    In contrast, deterministic theoretical models \cite{dank07,kaou09a,faru10,bibe11,zhao11,yazd12} taking next-order hydrodynamic terms or harmonics with $l>2$ into account argue that a third parameter, e.g., the excess area $\Delta$, is necessary. In particular, in pure shear flow, the reduced viscosity contrast $\Lambda$ at which the TT-TR/TB transition happens should strongly depend on $\Delta$ \cite{faru10,bibe11,zhao11}. Only for $\lambda=1$ do the phase diagrams for various $\Delta$ plotted as function of $S$ and $\Lambda$ overlap (see Fig. 4 of \cite{faru10}). This theoretical dependence on three parameters is in strong discrepancy with experimental results. The experimental phase diagrams in pure shear flow \cite{desc09}, where $\omega/s=1$ and $\lambda$ is varied, and in general linear flow \cite{desc09a}, where $\lambda=1$ and $\omega/s$ is varied, are both well described in terms of $S$ and $\Lambda$, see Fig. \ref{fig_pd}. Even if different values of $\Delta$ are distinguished, the scaling still holds \cite{zabu11}. Deterministic simulations suggest that experimental motions might be wrongly classified due to long transients \cite{bibe11,zhao11,yazd12}, but these are not observed in long experiments \cite{zabu11,leva12a}. Here again, thermal fluctuations have a great impact on the theoretical phase diagram \cite{abre13} and may be crucial to resolve the discrepancy.
    
    At large flow rates, several additional dynamical features are predicted, yet to be observed experimentally. An analytical model \cite{lebe08} and numerical simulations \cite{bibe11,zhao11} predict a three-dimensional motion called kayaking (or spinning), in which the vesicle precesses like a spinning top. For high excess areas corresponding to that of a RBC ($\Delta=5$), a recent analytical work predicts the existence of some peculiar deformation modes at flow parameters corresponding to VB/TR \cite{faru12b}. However, it is not clear whether these shapes subsist with thermal fluctuations. Moreover, shape deformations observed in \cite{yazd12,leva14} include higher-order harmonics, which is due to the wrinkling instability discussed in section \ref{sec_el2}. At even higher shear rates, inertial effects become important. For instance, simulations \cite{sala12,laad12,kim12,luo13} show that inertia inhibits tumbling. Experiments on vesicles exposed to acoustic streaming near a bubble seem to indicate that membrane rupture occurs at very large shear stresses \cite{marm08}. Such membrane rupture has, however, not yet been observed in linear flows at large flow rates.
  \end{subsection}

  \begin{subsection}{Stationary elongational flow}
    \label{sec_el}
    \begin{figure}
      \centering
      \includegraphics[width=0.9\linewidth]{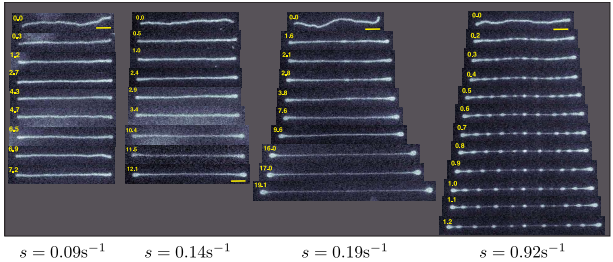}
      \caption{Stretching of a tubular vesicle at different flow strengths. The scale bar is 10 $\mu m$ and the numbers are the renormalized time $t^*=st$ (from \cite{kant08a}).}
      \label{fig_stretch}
    \end{figure}
    Setting $\omega=0$ in Eq. \eqref{linflow}, one obtains an elongational flow which stretches the vesicle along the $(y=x)$-axis. In contrast to droplets and elastic capsules, a vesicle deforms in such a flow only if it has a positive excess area $\Delta>0$, since its area and volume are constant and a sphere is the only possible shape for $\Delta=0$. For small excess areas, the vesicle deforms into an ellipsoid aligned with the flow, the deformation reaching a plateau at large elongation rates due to membrane incompressibility \cite{kant07,turi08,zhao13}. 
    
    For large excess areas, a transition similar to the coil-stretch transition of flexible polymers \cite{perk97} is found, see Fig. \ref{fig_stretch}. Under a critical elongation rate, floppy tubular vesicles are only slightly stretched by the flow. Above the critical rate, they evolve into dumbbells connected by a thin tether whose length grows with the magnitude of the flow. At even higher flow rates, the tether becomes unstable and transient pearling configurations with different number of beads are observed until a stationary stretched dumbbell shape is established \cite{kant08a}. This phenomenon resembles somewhat the laser-induced pearling instability of tubular vesicles \cite{barz94,nels95}. In flow, the tubule-to-dumbbell transition is a continuous one accompanied by a slowing down of the dynamics and enhanced fluctuations close to the critical elongation rate, like for the coil-stretch transition of polymers \cite{gera08}. This transition was observed recently in numerical simulations which showed asymmetric dumbbells and dynamical tether extension \cite{zhao13}. Such membrane nanotubes are used for intercellular communication or for directing the spread of viruses \cite{sowi08}.
  \end{subsection}

  \begin{subsection}{Time-dependent elongational flow}
    \label{sec_el2}
    \begin{figure}
      \centering
      \includegraphics[width=0.9\linewidth]{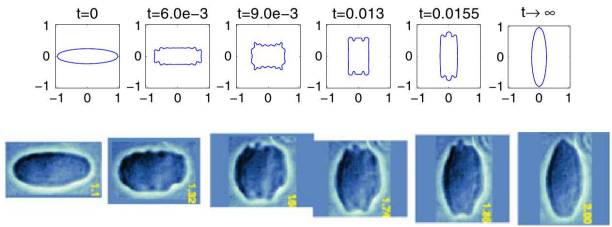}
      \caption{Wrinkling phenomenon in two-dimensional simulations (top, from \cite{liu13}) compared with experimental shapes (bottom, from \cite{kant07}).}
      \label{fig_wrink}
    \end{figure}
    If the direction of the elongational flow is suddenly reversed, the transient phenomenon of vesicle wrinkling is observed \cite{kant07,turi08,liu13}. The vesicle, which is initially stretched by the flow, is temporarily compressed after the flow reversal. This transient compression leads to an effective negative surface tension which amplifies harmonics with $2<l<l_{max}$ where the highest mode $l_{max}$ grows with the flow strength \cite{turi08}. Recent two-dimensional simulations \cite{liu13} have shown qualitatively similar wrinkles, see Fig. \ref{fig_wrink}, even though, contrary to experiments \cite{kant06}, odd harmonics are not observed due to the absence of thermal noise. This behavior has clearly to be distinguished from the steady-state wrinkling of elastic capsules in elongation flow, which is due to the elasticity of the capsule membrane \cite{fink06}, whereas vesicles are, in good approximation, inelastic.

    \begin{figure}
      \centering
      \includegraphics[width=0.9\linewidth]{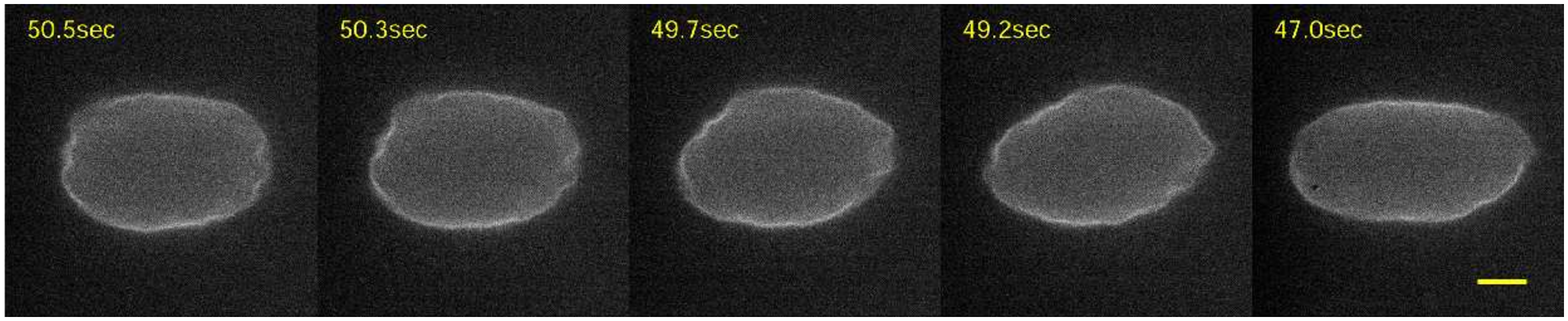}
      \includegraphics[width=0.9\linewidth]{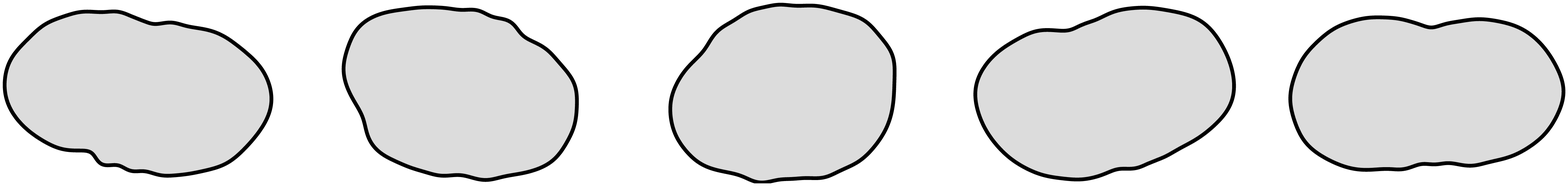}
      \caption{Wrinkling of a vesicle in TR motion at high $S$ in experiments (top) and stochastic simulations (bottom) (from \cite{leva14}).}
      \label{fig_wr}
    \end{figure}
    This phenomenon was recently observed experimentally in steady linear flow for trembling vesicles \cite{leva14}. Due to the quasi-periodic oscillation of its orientation, a vesicle in TR motion is alternately stretched and compressed by the elongational part of the flow. Therefore, the surface tension becomes quasi-periodically negative and higher-order harmonics are excited, see Fig. \ref{fig_wr}. Contrary to the previous case, high-order modes do not necessarily decay fully until the next compression period, making the dynamics very complex. However, the mean wave-number of shape deformation depends on the flow strength in a similar fashion as in the case of time-dependent elongational flow \cite{leva14}.
  \end{subsection}

\end{section}

\begin{section}{Walls, capillaries, and rheology}
  \begin{figure}
    \centering
    \includegraphics[width=0.5\linewidth]{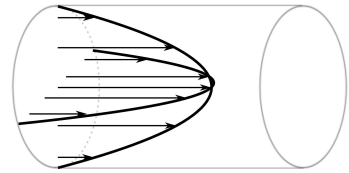}
    \caption{Poiseuille flow in a cylindrical channel. The length of the arrows gives the flow velocity at the entrance of the channel.}
    \label{fig_pois}
  \end{figure}
  The study of vesicle dynamics is often cited as a first step to model blood flow \cite{vlah13}. In small capillaries, blood velocity has a parabolic profile also known as Poiseuille flow, see Fig. \ref{fig_pois}. The maximal velocity, the degree of confinement, and the vesicle properties control the dynamics. Close to the walls, a Poiseuille flow can be approximated by a linear shear flow. We will thus first describe the dynamics of a single vesicle close to a wall in shear flow, before turning to Poiseuille flows, vesicle interaction, and rheology.

  \begin{subsection}{Influence of walls in shear flow}
  \label{sec_lift}
    \begin{figure}
      \centering
      \includegraphics[width=0.9\linewidth]{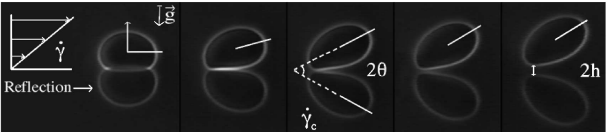}
      \caption{Deformation and lift of a vesicle due to a shear flow close to a wall (from \cite{abka02a}). The shear rate is increased from left to right. Above the critical shear rate $\dot{\gamma}_c$, the lift force is large enough to counteract gravity and the vesicle remains at constant height $h$ from the wall.}
      \label{fig_lift}
    \end{figure}
    In linear shear close to a wall, a spherical object would move along with the flow while remaining at a constant distance from the wall because of the reversibility of the Stokes equations. One expects a transverse migration only for high Reynolds number due to the Magnus effect. In contrast, an object with broken fore-aft symmetry, such as a tank-treading vesicle with positive inclination angle, experiences a force pushing it away from the wall. In particular, a deformable vesicle sticking to a wall due to gravitational or adhesion forces is lifted by a strong enough flow, see Fig. \ref{fig_lift}. By increasing the shear rate, the vesicle first deforms into an asymmetric shape and acquires a positive tilt, which increases with the shear rate. Above a critical shear rate $\dot{\gamma_c}$, the vesicle is lifted and remains at a constant distance $h$ from the wall where gravitation compensates the hydrodynamic lift \cite{seif99a,cant99,lorz00,suku01,abka02a,abka05}.\footnote{The depinning dynamics of a settled vesicle have also been studied under an axisymmetric suction flow \cite{chat09}.} In the absence of gravitation, the vesicle drifts away from the wall with constant velocity \cite{call08}. Far from the wall, i.e., for $h>R_0$, the lift force scales with $\dot{\gamma}/h^2$, as predicted theoretically \cite{olla97,olla97a} and observed in experiments \cite{call08} and numerical simulations \cite{suku01,me09,zhao11b}. Closer to the wall, the $\dot{\gamma}/h$ scaling observed in experiments \cite{abka02a,abka05} is reproduced by recent numerical simulations \cite{zhao11b}. Moreover, the lift force grows with increasing excess area $\Delta$ \cite{olla97,abka02a,abka05,me09}. This lift force may be responsible for the cell-free layers near the walls of blood vessels known as the F{\aa}hraeus-Lindqvist effect \cite{olla99,abka05,lamu13}.

    For TB vesicles, the lift force was predicted to vanish since the motion is somewhat symmetric \cite{olla00}. However, experiments with RBCs show that a lift force subsists in TB motion due to asymmetric deformations \cite{gran13}. If the center of the vesicle is closer to the wall than its radius, TB cannot happen. Even far enough from the wall, the TT-TB transition occurs at a higher critical viscosity contrast $\lambda_c$ than in an unbounded flow because of the increased pressure between the wall and lower tip of the vesicle \cite{me09,zhao11b}. A similar increase of $\lambda_c$ is also observed in a shear flow confined between two walls, even if these walls are separated by a distance much larger than the vesicle radius \cite{beau04,kaou11,kaou12}. More specifically, the closer the two walls are, the higher $\lambda_c$ becomes. This factor might contribute to the shear thinning of blood in small capillaries \cite{kaou11,kaou12}. 
  \end{subsection}
  
  \begin{subsection}{Capillary flows}
    \begin{figure}
      \centering
      \includegraphics[width=0.5\linewidth]{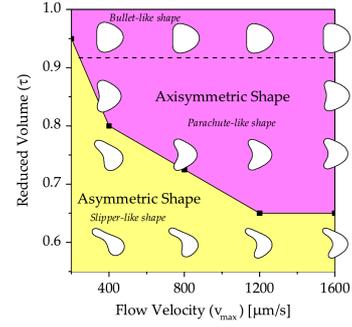}
      \caption{Shapes of a vesicle in a two-dimensional unbounded Poiseuille flow as a function of the reduced volume and the maximal flow velocity (from \cite{kaou09}).}
      \label{fig_parab}
    \end{figure}
    In a Poiseuille flow, vesicles migrate towards the tip of the flow where the shear rate is minimal \cite{coup08,kaou08,dank09}. \footnote{If the flow were curved, however, the migration would be towards regions of higher curvature, i.e., higher shear rates \cite{ghig11}.} For low excess areas $\Delta$, vesicles adopt a shape which is symmetric relative to the the axis of symmetry of the flow. Contrary to the case of linear flow, spherical harmonics with $l=3$ are also excited to lowest order \cite{olla00}. The two possible axisymmetric shapes are the bullet-like shape, with a convex rear end, and the parachute-like shape, with a concave rear end \cite{vitk04,dank09}. For larger excess areas, these axisymmetric shapes become unstable and models predict an asymmetric off-centered slipper shape, which has already been observed for RBCs \cite{nogu05a,nogu05b,kaou09,faru11,mcwh11}. Fig. \ref{fig_parab} shows the theoretical shape diagram of vesicles in two-dimensional unbounded Poiseuille flow as a function of the flow velocity and the reduced volume $\tau\equiv(1+\Delta/4\pi)^{-3/2}$, in which the three possible shapes are present. For slower flow rates as presented in this diagram, discoidal, prolate, or stomatocyte shapes are also predicted \cite{nogu05a,nogu05b,mcwh11}.
    
    The previous results hold in unbounded flows. For small capillaries, confinement induces new effects, for example cross-streamline migration due to the wall lift force, see Section \ref{sec_lift}. Vesicles migrate towards the center due to the walls and to the flow profile, with a migration velocity proportional to $\dot{\gamma}(h)/h$, where $\dot{\gamma}(h)$ is the local shear rate at a distance $h$ from the wall \cite{coup08}. Once at the center of the channel, vesicles move with the flow. Their velocity is then almost equal to the maximal velocity of the flow for low confinement, but much lower for strong confinement \cite{brui96,vitk04}. In asymmetric channels, recent experiments show the existence of a croissant-like shape, for which the rear end is convex in one direction but concave in the perpendicular one \cite{coup12}. Such three-dimensional effects cannot be captured by two-dimensional models, which, however, reproduce many experimentally observed features and make predictions that have not been tested yet in experiments \cite{kaou11a}. A good agreement between models and experiments has been obtained for vesicles and RBCs flowing in micro-channels of oscillating width \cite{nogu10d,brau11}. In such channels, orientation and shape oscillations occur and asymmetric slipper-like shapes are observed. 
  \end{subsection}
  
  \begin{subsection}{Hydrodynamic interactions and rheology}
    \begin{figure}
      \centering
      \includegraphics[width=0.9\linewidth]{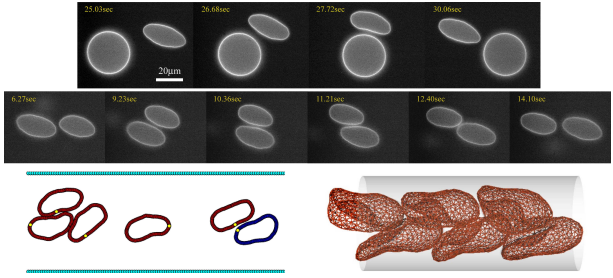}
      \caption{Top: TT dynamics of a vesicle altered by a presence of an additional vesicle (from \cite{leva12}) Bottom: Simulations of interacting vesicles in two (left, from \cite{lamu13}) and three dimensions (right, from \cite{mcwh09}).}
      \label{fig_inter}
    \end{figure}
    Using the results obtained for quasi-spherical vesicles (see Sect. \ref{sect_lin}), the effective viscosity of a dilute suspension of vesicles in shear flow can be calculated \cite{dank07,verg08}. For TT vesicles, the analytical expression 
    \begin{equation}
      \frac{\eta_{eff}-\eta_o}{\eta_o\phi}=\frac{5}{2}-\frac{\Delta}{16\pi}(23\lambda+32)
      \label{eq_visceff}
    \end{equation}
    has been obtained \cite{dank07}, where $\eta_{eff}$ is the effective viscosity of the suspension and $\phi$ the volume fraction of vesicles. The term $5/2$ corresponds to Einstein's correction for hard spheres. In these models, the effective viscosity first decreases with the viscosity contrast and reaches a minimum at the TT-TB transition ($\lambda_c>1$), then grows with $\lambda$ in the TB regime \cite{dank07,rahi10,ghig10,thie13,zhao13a}.  While such a cusp has been observed in experiments with vesicles and RBCs \cite{vitk08}, other experiments show that for $\lambda<1$, $\eta_{eff}$ rather increases with $\lambda$ \cite{kant08}. This discrepancy may come from the fact that rheological measurements of vesicle suspensions are complicated due to the wide dispersion of physical parameters of the constituents, such as size and excess area, and the inability to measure them directly. Therefore, individual objects in solution do not necessarily undergo the same dynamics, making the micro-macro relations much more complicated. Moreover, long-range hydrodynamic interactions between vesicles cause strong shape fluctuations even at relatively low volume fractions \cite{kant08}. A way to determine the volume fraction at which a vesicle solution can be considered as dilute is to look first at the interaction between two vesicles \cite{rama10,le11,gire12,leva12,lamu13}. Recent experiments show that hydrodynamic interactions are important for volume fractions larger than $\phi=0.08-0.13$, a result confirmed by the measurement of the back-reaction of a single vesicle on the ambient flow \cite{leva12}. These interactions strongly influence the orientation and shape of the vesicles, in particular during crossing motions \cite{leva12,lamu13,zhao13a}, see Fig. \ref{fig_inter}. In addition, vesicles repel each other if they are in the same shear plane but hydrodynamic attraction can occur otherwise \cite{gire12}.

    For concentrated solutions in shear flow, shear-thinning happens because the cell-free layer at the walls grows with the shear rate in TT \cite{freu11,lamu13}. In capillary flows, the effect of walls, shear gradient, and hydrodynamic interactions add up to produce non-trivial vesicle migration dynamics \cite{faru13}. Vesicles flowing at the center of the channel form clusters due to hydrodynamic interactions if they come close enough to each other \cite{ghig12} similar to RBC clusters observed experimentally \cite{toma12}. A transition between single files of parachute-shaped vesicles and a zigzag arrangement of slippers (bottom right of Fig. \ref{fig_inter}) is predicted with increasing confinement and happens continuously due to thermal fluctuations \cite{mcwh09,mcwh11,mcwh12}.
  \end{subsection}
\end{section}

\begin{section}{Outlook}
  The dynamics of single vesicles in steady viscous flow is by now well understood. Many experimental observations have been successfully explained by models and simulations, even though the quantitative agreement between theoretical predictions and experimental results has to be improved, for example with the inclusion of thermal fluctuations and inertial effects in numerical models. Moreover, there are still a lot of questions to be answered as far as more complex setups are concerned. The next points are only an outline of some of the future perspectives in the field. 
  
  \begin{subsection}{Relaxation of stretched vesicles}
    A vesicle deformed by an external flow will relax back to its equilibrium shape once the flow is turned off. For vesicles relaxing from dumbbell to tubular shapes in elongational flow as described in Section \ref{sec_el}, relaxation times of several seconds have been found \cite{kant08a}. The same order of magnitude has been found with vesicles deformed by a strong uniform flow while trapped by optical tweezers \cite{foo04} or directly deformed by optical tweezers \cite{zhou11}. This relaxation time also depends on the initial tension of the vesicle, as shown by experiments with vesicles point-attached to a solid substrate or to a moving particle \cite{ross03}. In any case, the experimental relaxation times are an order of magnitude smaller than estimated from a model for quasi-spherical vesicles \cite{lebe07,lebe08}, showing the need for a better modeling of these processes.
  \end{subsection}
  
  \begin{subsection}{Time-dependent flow}
    In blood flow, RBCs do not experience a steady flow but rather an oscillatory one. Experiments on RBCs in oscillatory shear flow have been done \cite{naka90,wata06,dupi10}, and among other features chaotic motions have been observed \cite{dupi10}. For vesicles, however, there are not so many studies in oscillatory shear flow. One experiment shows that fluctuations are suppressed by such a flow \cite{fa04}. Numerical simulations for vesicles and capsules show complex dynamical regimes, in particular a delay between vesicle and flow dynamics and symmetry-breaking through thermal noise \cite{nogu10,nogu10b,zhao11a}. For dilute suspensions of vesicles, the effective viscosity is expected to show resonances \cite{faru12} similarly to what happens for elastic capsules \cite{kess09}. These questions and many others still have to be investigated in time-dependent flows.
  \end{subsection}

  \begin{subsection}{Dynamics of compound vesicles in linear flow}
    \begin{figure}
      \centering
      \includegraphics[width=0.6\linewidth]{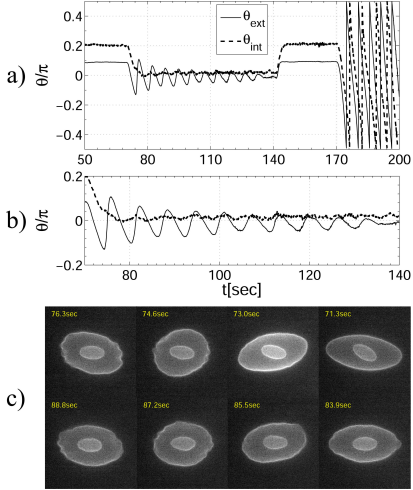}
      \caption{(a) Time evolution of the orientation of the external vesicle $\theta_{ext}$ and the internal one $\theta_{int}$ in consecutively TT, TR, TT and TU regimes. (b) Magnification of (a) showing the swinging regime. (c) Snapshots of two subsequent cycles in TR dynamics with clearly visible wrinkles.}
      \label{fig_comp}
    \end{figure}
    Compound vesicles are GUVs with another object inside. They are used to model the dynamics of cells having a complex internal structure such as white blood cells (WBCs). Recent simulations considering a solid inner particle of either spherical or elliptical shape show that new dynamical features in shear flow appear due to the hydrodynamic interaction between the inclusion and the membrane \cite{veer11}. For example, the transition from TT to TB can occur in the absence of any viscosity contrast but at some critical value of the filling factor. Moreover, a vesicle can swing if the enclosed particle is non-spherical. A compound vesicle including another vesicle instead of a solid particle (referred to as bilamellar vesicle) demonstrates even more complex behavior \cite{kaou13}. Experimental observations of such a vesicle in linear flow were realized in a four-roll mill device. It is found that while a compound vesicle can undergo the same TT, TR, and TU regimes as a GUVs, a new swinging motion of the inner vesicle is found in TR in accord with simulations \cite{leva13}. In addition, the inner and outer vesicles can exist simultaneously in different dynamical regimes and be either synchronized or unsynchronized depending on the filling factor $\phi$, see Fig. \ref{fig_comp}. Multilamellar vesicles also deform and tank-tread in a complex fashion \cite{pomm13}. Whether these complicated models can mimic the dynamical behavior of WBCs or whether one needs to take an additional non-Newtonian nature of the WBC  and its nucleus (consisting of fibers, organelles, cytoplasm, etc...) into account remains an open question.
  \end{subsection}

  \begin{subsection}{Role of spontaneous curvature and area-difference elasticity}
    Essentially all theoretical work for vesicles in flow has used the minimal model for curvature energy \eqref{helf}. Shape transformations in equilibrium are known to depend on a finer description with spontaneous curvature and/or area-difference elasticity \cite{seif97}. These effects should also be investigated for vesicles in flow, in particular since work in other non-equilibrium situations \cite{sens04,fran09,rahi12} has shown their relevance. For instance, such bilayer effects might be important to explain the membrane budding observed in experiments in flow \cite{kant07,zabu11}.
  \end{subsection}
  
  \begin{subsection}{Multicomponent flows}
     Numerical simulations  \cite{bagc07,fedo11,freu11,tan12} and experiments \cite{gran13} with only RBCs show shear-induced diffusion as well as the F{\aa}hraeus-Lindqvist effect, i.e., the cell-free layer at the walls associated with a decrease in viscosity with the size of capillaries. Blood is, however, not only composed of RBCs but also of WBCs. In blood flow, it is observed that RBCs migrate away from the walls while WBCs migrate towards them, a phenomenon known as margination \cite{seco13}. Simulations of multi-component flows consisting of stiff and soft particles show that margination is probably due to this difference of stiffness \cite{freu07,fedo12,kuma12,kuma12a}. Moreover, recent experiments use the wall-induced lift force discussed in Sect. \ref{sec_lift} to efficiently separate cancer cells from RBCs \cite{geis13}. One can therefore wonder whether such phenomena could be observed in suspensions of GUVs and compound vesicles. 
  \end{subsection}

  \begin{subsection}{Other setups and related objects}
    Vesicles have also been studied under other non-equilibrium conditions. They move and deform under the influence of gravity \cite{krau95,veer09,huan11,boed12,boed13,suar13}, electromagnetic fields \cite{kumm91,sens02,rist10,sali12,seiw13,zhan13}, mechanical forces \cite{ross03,zhou11,chen12,guve13}, or pH-gradients \cite{nawa13}. Further investigation of such phenomena will constitute a basis for modeling the individual and collective behavior of complex biological cells, as well as provide tools to understand the physics of microscopic non-linear systems.

    Finally, many objects similar to vesicles have been studied under comparable conditions. Floppy vesicles with very low bending rigidity behave similarly to polymers \cite{gomp93,gomp95,krol95,shah98}. Ring polymers also tank-tread and tumble \cite{chen13}. As opposed to vesicles, liquid droplets \cite{rall84,ston94} and capsules with visco-elastic membranes \cite{pozr03a,bart09,fink11,bart11,gao11,le11,gao12,yazd13,wang13} respond in a linear fashion to shear flow since they can be stretched. Red blood cells also possess an incompressible bilayer membrane but are much less deformable than vesicles due to their cytoskeleton \cite{abka08,vlah09,basu11,misb12,vlah13}. They exhibit not only TT and TB regimes, but also many other motions due to their elasticity \cite{dupi12}. Moreover, their membrane is much more complex than that of vesicles and they are sensitive to changes in ATP-concentration \cite{fors11,brau12}. The investigation of the dynamics of such objects will help understand to which extent vesicle properties can be generalized to other physical systems. 
  \end{subsection}
\end{section}

\begin{section}{Acknowledgments}
  U. S. dedicates this review to Wolfgang Helfrich on the occasion of his 80\textsuperscript{th} birthday. We acknowledge financial support from the German-Israeli Foundation.
\end{section}

\bibliographystyle{unsrt}
\bibliography{fluves.bib}

\end{document}